\documentclass[11pt, twoside, a4paper, final]{article}
\usepackage{graphicx, epstopdf, url, tabularx, array, booktabs, color}
\usepackage[usenames,dvipsnames]{xcolor}
\usepackage[margin=2.5cm]{geometry}
\usepackage[utf8]{inputenc}

\date{December 22, 2016}

\begin{document}
\title{Security-related Research in Ubiquitous Computing -- Results of a Systematic Literature Review}
\author{Ema Ku\v{s}en \and Mark Strembeck}
\maketitle
\begin{abstract}
  In an endeavor to reach the vision of ubiquitous computing where
  users are able to use pervasive services without spatial and
  temporal constraints, we are witnessing a fast growing number of
  mobile and sensor-enhanced devices becoming available. However, in
  order to take full advantage of the numerous benefits offered by
  novel mobile devices and services, we must address the related
  security issues. In this paper, we present results of a systematic
  literature review (SLR) on security-related topics in ubiquitous
  computing environments. In our study, we found
  5165 scientific contributions published
  between 2003 and 2015. We applied a systematic procedure to identify
  the threats, vulnerabilities, attacks, as well as corresponding defense
  mechanisms that are discussed in those publications. While this
  paper mainly discusses the results of our study, the corresponding
  SLR protocol which provides all details of the SLR is also publicly available 
	for download.
\end{abstract}


\section{Introduction}

About two decades ago, Mark Weiser
\cite{Weiser91} coined the term
``ubiquitous computing''. He envisioned a world enhanced with
technologies that disappear into the background and become invisible
to the user.  Once only a fiction, the proliferation of smart wearable
devices and sensor-enhanced mobile phones on the market worldwide,
makes it easy to imagine a not so distant future in which the
unobtrusive technologies of ubiquitous computing will be an
indispensable part of our daily life. Apart from influencing the way
we communicate and interact with our environment, these new
technologies are slowly retracting into the user's most private
spheres, homes and hobbies. Based on the IHS iSuppli's estimate, in
just three years since the beginning of our study (2013) wearable fitness gadgets will increase in
their shipment by 12.4 million units
\footnote{http://www.statista.com/statistics/259597/fitness-gadget-shipments-forecast/}.
Such a future sees an increasing number of people using gadgets to
learn more, not only about the air temperature or the nearest coffee
shop, but also about their heartbeat rate, quality of sleep, stress
level, and other potentially sensitive information.

Yet, the progress in technology is often accompanied by adversaries
finding new means to break into a system and steal or corrupt
sensitive information. Ubiquitous computing is no different. Recently,
news such as ``Google Glass doesn't have a privacy problem. You
do''\footnote{http://time.com/103510/google-glass-privacy-foregrounding/}
have raised the attention to rethink which information we are willing
to share and by which means. Threats such as modifying sensor readings
in a wearable medical device or tracking a user's whereabouts have
already been recognized as serious challenges that have to be resolved
if pervasive computing is to be fully adopted into the society.

In this paper, we present results of a systematic literature review
(SLR) on scientific publications that address security-related topics
in ubiquitous computing environments.  In particular, this paper
provides supplemental analyses that have not been included in
\cite{ubisec-slr-2016}.

\section{Research method}

With its origins in the discipline of evidence-based medicine, SLR as a method has been
adopted to the software engineering domain since (at least) 2004
\cite{Kitchenham2009} to provide an evidence-based overview of the
existing body of knowledge found in the scientific literature
\cite{Kitchenham2013}. In contrast to other review methods, SLR tries
to combine the findings of all relevant studies to increase the
statistical significance and reliability of the obtained results and
ensure their transferability to other settings \cite{Chiappelli10}.
Although variations in conducting an SLR exist, the review typically
includes three phases:

\begin{enumerate}
\item \textit{Planning the review:} The SLR begins by defining a
  research question and specifying the search procedure, inclusion
  criteria against which the papers found during the SLR will be
  evaluated (see Table \ref{tab:exclusion}), criteria used to assess papers for their quality (see Table \ref{tab:exclusion}), and   information to be recorded during the data extraction procedure.
  From the beginning, all steps of an SLR are recorded in a
  corresponding \emph{SLR protocol}. This protocol is then used
  throughout the second phase of the review (see below) to ensure rigor. Please refer to our research protocol
  for more details at: http://epub.wu.ac.at/4826/.
	
\item \textit{Conducting the review:} The search procedure usually
  consists of automated steps as well as steps that are conducted
  manually. Manual steps include screening the papers in the selected
  journals, conference proceedings, and other sources, while the
  automated steps apply carefully designed search strings to
  scientific databases and automatically collect the search results. Once the
  initial pool of papers has been obtained, the SLR proceeds with
  screening of the papers and exclusion of the ones that do not
  satisfy the inclusion criteria.  Moreover, each paper that survives
  the filtering process is assessed for its quality according to the
  predefined set of criteria described in the protocol. The final step
  of the second phase includes extraction of the information found in
  the papers and summarizing the evidence.
\item \textit{Reporting on the results:} The SLR finishes by
  interpretation and discussion of the obtained results.
\end{enumerate}

\textbf{Notes on our experience.} Conducting an SLR has proven to be quite a lengthy process. We started with a protocol development in December 2013 and continued with a search for papers in late January 2014. We restricted the search to papers written in English that were published within 2003-2013 and indexed in 5 scientific databases (ACM Digital Library, IEEE Xplore, Science Direct, Springer Link, Wiley Digital Library). After performing the first search round, we refined the list of key terms based on the results of our pilot search. For example, the key term \textit{safety}, which was included in our list in the early stage, resulted in a large number of articles that were out of the scope of our SLR, such as publications on homeland security and child safety. The final list of key terms included two groups related to:
\begin{enumerate}
	\item \textbf{ubiquitous computing} (ubiquitous computing, pervasive computing, wearable computing, body area network, mobile computing), and
	\item \textbf{security} (security, confidentiality, authentication, access control, non-repudiation, audit, integrity, authenticity of data, availability, accountability, privacy, trust).
\end{enumerate}

\begin{figure}[ht]
	\centering
		\includegraphics[scale=0.6]{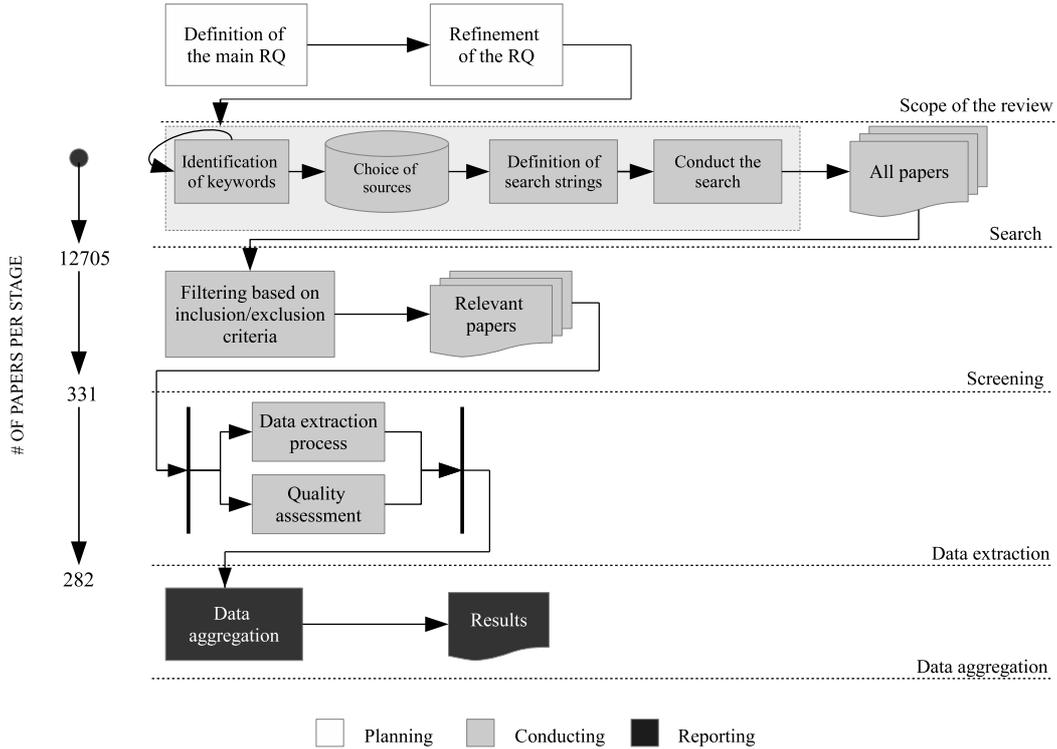}
	\caption{Overview of our SLR procedure.}
	\label{fig:procedure}
\end{figure}

Initially (in February 2014), we found 4492 papers that were
potentially to be included in our SLR. Over the course of two months,
we used our predefined exclusion/inclusion criteria
(see Table \ref{tab:exclusion}) to filter
papers that were irrelevant for our SLR.  In
  January 2015, we repeated the search to obtain papers published in
  2014 and included them in our SLR, which gave us a total number of
  5109 papers that have undergone the filtering procedure. In total
  4872 papers were excluded from our SLR based on the criteria
  presented in Table \ref{tab:exclusion}. The remaining 237 papers
  were taken to the next stage, in which we performed data extraction
  and quality assessment procedures in parallel (see Figure
  \ref{fig:procedure}). We based our quality assessment (see Table
  \ref{tab:exclusion}) on the suggestions by \cite{Kitchenham2013,Afzal2009,Radjenovic13} and used a three-point scale
  with \textit{Yes} (1), \textit{No} (0), and \textit{To some extent}
  (0.5) as possible scores for each question. In total, 14 papers were
  excluded based on a low quality assessment score, which left us with
  223 papers that were analyzed in detail to obtain results for our
  SLR.

\begin{table}[hbp]
\scriptsize
	\begin{tabularx}\textwidth{p{1cm} p{14cm}}
		\toprule
			\textbf{ID} & \textbf{Exclusion criteria}\\
			\midrule
			E01 & Summaries of workshops and tutorials, title pages, editorials, and extended abstracts as they do not provide sufficient information to the objective of our review. \\
			E02 & Workshop papers as they report on a study in its early stage.\\
			E03 & Posters, as they do not provide enough information for the purpose of our review.\\
			E04 & Books and PhD theses, as they are beyond the scope of this review.\\
			E05 & Double entries. If an extended journal article is found, it will be chosen over the conference article. If a more recent paper is found, it will be chosen over its preceding paper.\\
			E06 & Papers whose focus is not put on security-related research in ubiquitous computing, i.e. papers that just mentioned security in their abstracts as one of the issues. \\
			E07 & Opinion papers and discussion papers that do not propose a solution.\\
			E08 & Any paper whose full text is not accessible.\\
		  E09 & Papers not written in English.\\
			E10 & Papers with a low quality assessment score. \\
			\midrule
			\textbf{ID} & \textbf{Inclusion criteria}\\
			\midrule
			I01 & Full version of journal and conference articles that report on, discuss, or investigate security issues in ubiquitous computing. \\       
	    I02 & Papers that propose a solution to the identified security issue.\\
			I03 & Papers written in English.\\
			I04 & Papers published since 2003.\\
		\bottomrule
		\textbf{ID} & \textbf{Quality assessment} \\
		\midrule
		  QA1 & Is the paper based on research?   \\
			QA2 & Is there a clear statement of the aim?   \\
			QA3 & Is there an adequate description of the context in which the research was carried out? \\
			QA4 & Did the paper make a review of previous research of the topic?  \\
			QA5 & Is the methodology described adequately?  \\
			QA6 & Is there a clear statement of the findings?  \\
			QA7 & Did the paper discuss future work?  \\
			\bottomrule
		\end{tabularx}
	\caption{Exclusion/inclusion criteria and quality assessment.}
	\label{tab:exclusion}
\end{table}

\section{Security challenges in ubiquitous computing: vulnerabilities, threats, and attacks}

As a part of our SLR, we have identified the vulnerabilities, threats, and attacks that are frequently addressed in scientific publications on ubiquitous computing environments.

\subsection{Vulnerabilities}

\begin{figure}[ht]
	\centering
		\includegraphics[scale=0.9]{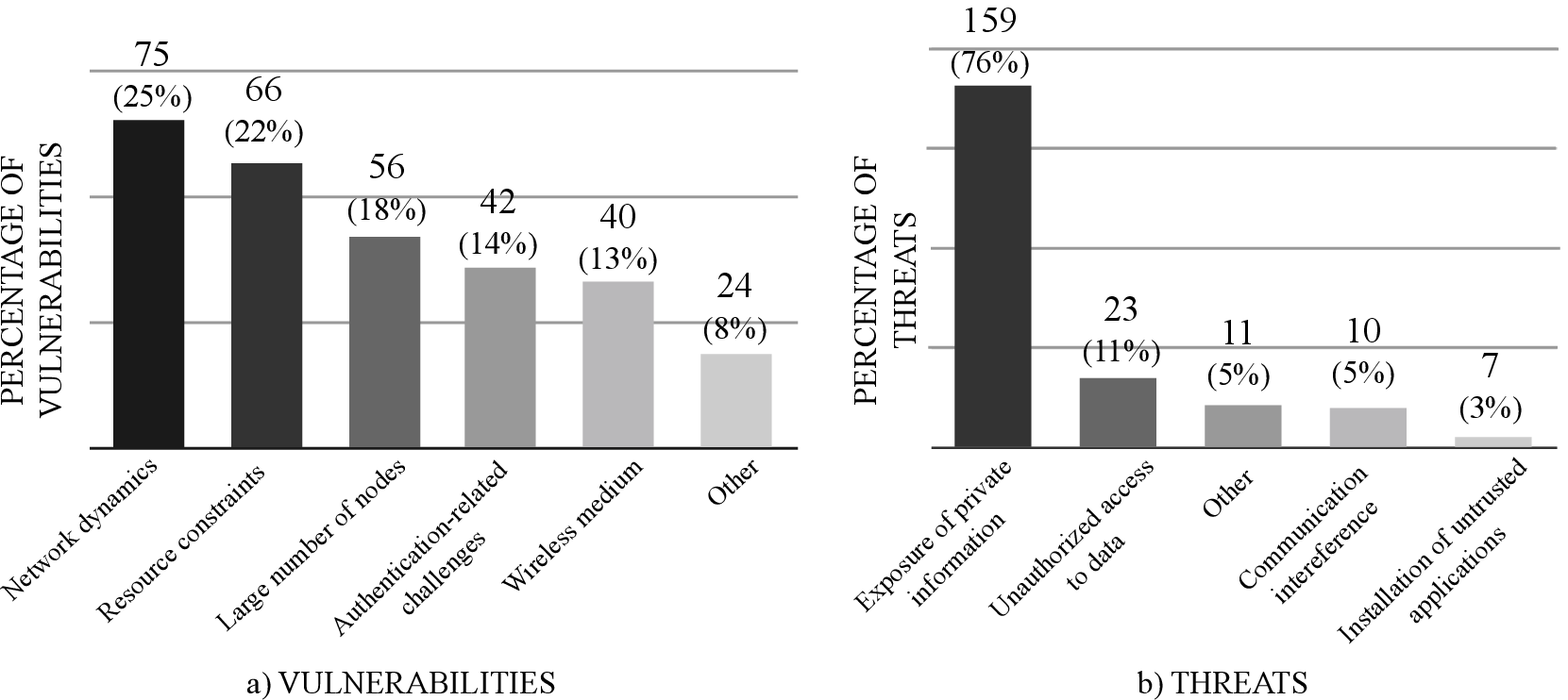}
	\caption{Vulnerabilities and threats in ubiquitous computing.}
	\label{fig:vulnerabilitiesThreats}
\end{figure}

\textbf{Network dynamics.} The results of our SLR (see Figure \ref{fig:vulnerabilitiesThreats}) show that the most addressed vulnerability is related to network dynamics. As mobile devices may join or leave the network at any given time, it is important for a network to self-configure. Such a scenario is characteristic for its absence of a fixed infrastructure and a lack of a central server, central authority, and centralized trusted third party (see, e.g.\cite{Zhang03}).

\textbf{Large number of nodes.} The second most commonly addressed challenge results from the large number of nodes that engage in network communications. Since some nodes may act selfishly (refuse to
forward packets to other nodes), maliciously (seek to damage network operations) (see, e.g., \cite{Luo04}) or show signs of a dynamic personality (behave strategically in a way that best benefits them)
(see, e.g., \cite{Das12}), challenges occur in trust computation and management, as well as in the detection of ill-behaved nodes, which, if not handled carefully, may lead to a collapse of a whole network (see e.g., \cite{Ahamed09}).

\begin{figure}[ht]
	\centering
		\includegraphics[scale=0.5]{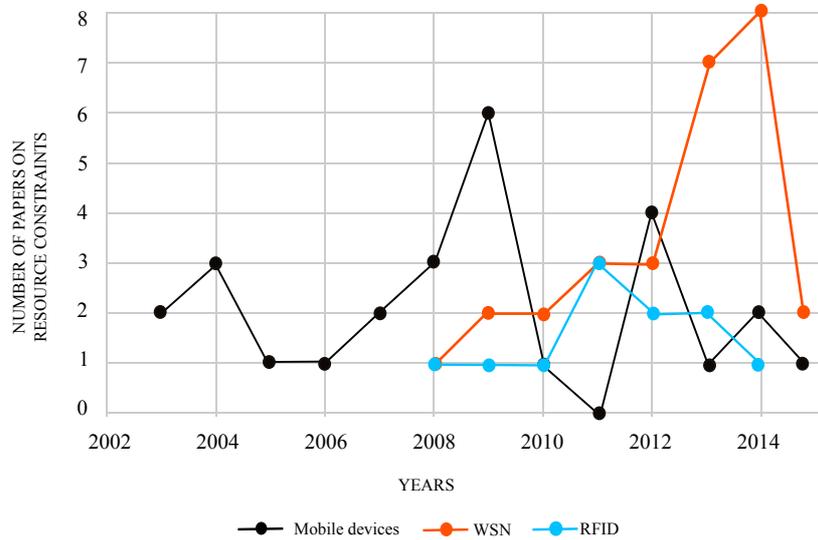}
	\caption{Resource constraints found in WSN, mobile devices, and RFID-based systems over the last decade.}
	\label{fig:resource-constraints}
\end{figure}

\textbf{Resource constraints.} Nodes of a ubiquitous computing environment are heterogeneous devices that differ in their processing power, battery-life, and communication capabilities. In ubiquitous
computing, it is challenging to ensure the availability of services and design security mechanisms that rely on complex computations due to the devices' resource constraints (addressed in 53 papers).  Our findings show that studies mainly address resource constraints of mobile devices, wireless sensor networks (WSN), and RFID-based systems (see Figure \ref{fig:resource-constraints}). The black line in Figure \ref{fig:resource-constraints} represents the number of studies that address resource constraints found in mobile devices. The results of our SLR indicate that the overall amount of studies that focus on resource constraints of smartphones has diminished over the last decade. On the other hand, the number of studies on resource constraints in sensor nodes rises, as shown by the red line. To a lesser amount, our SLR has also identified studies that address resource constraints in RFID-based systems, plotted with a blue line in Figure \ref{fig:resource-constraints}.
 
\textbf{Authentication-related challenges.} Our SLR shows that researchers are widely proposing adjustments to the existing authentication mechanisms in order to tailor them to the characteristics of ubiquitous computing. An important novel aspect of these mechanisms is unobtrusiveness. As envisioned by Mark Weiser, the goal of ubiquitous computing is to design environments in which people 
cease to be aware of the technologies surrounding them. Authentication mechanisms still do not entirely comply to this vision. For instance, although mobile phone locking mechanisms, such as passwords or pattern-drawing, are necessary to protect user's sensitive data from the unauthorized access, they still require attention from device owners, distracting them from their surroundings. Moreover, by owning more devices, users face challenges in remembering all the passwords. As a result, some users use the same password for multiple devices or store the list of passwords on their device, opening another way for an adversary to gain access to user's sensitive information. Moreover, some users entirely disable their locking mechanisms, leaving their devices vulnerable. In total, authentication-related challenges are addressed in 41 studies analyzed in our SLR.

\textbf{Other vulnerabilities.} Other vulnerabilities found in our SLR include lack of a prior knowledge of services, Bluetooth vulnerabilities, frequent change of a user's context, and installation of untrusted mobile apps.

\subsection{Threats}

As a part of our SLR, we have also identified threats that are addressed in scientific publications and grouped them into five categories (see Figure \ref{fig:vulnerabilitiesThreats}b).
The most frequently addressed threat is exposure of user's private information. Even though some information is provided accidentally by a user due to the lack of awareness, other can be recorded without a notification or user's consent. To better understand the privacy-related challenges, we have examined the corresponding papers in more depth.

\begin{figure}[ht]
	\centering
		\includegraphics[scale=0.7]{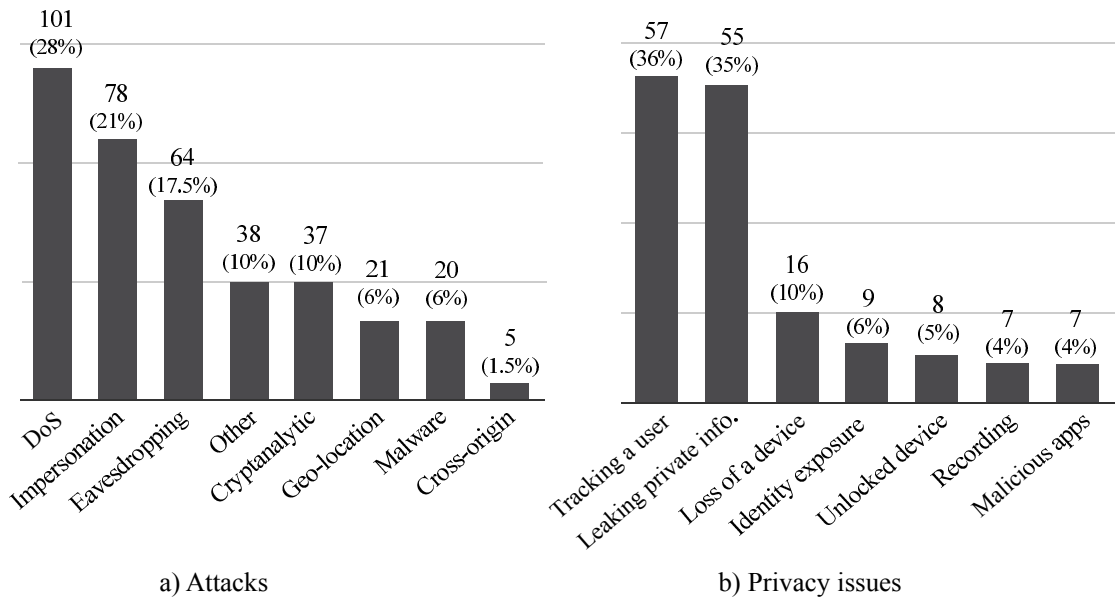}
	\caption{Attacks and privacy issues in ubiquitous computing.}
	\label{fig:attacks_privacy}
\end{figure}

As shown in Figure \ref{fig:attacks_privacy}b, the majority of the studies addressing privacy focus on the issue of attaining the current geographical position of a mobile device or user tracking (44,41\%), followed by leaking of private information while using mobile services (30\%). To a lesser amount, our SLR has also identified threats such as theft or loss of a mobile device, such as smartphones or smart watches (10\%), and leaving a user's device unlocked (6.5\%).  In addition to mobile devices, our SLR also found other technologies that store personal information, such as RFID tags found in access cards, badges, credit cards, passports, and driver's licenses. If these devices get lost or stolen, an adversary may not only get access to the owner's personal information, but also physically enter his/her home or office. Another threat identified in our SLR refers to recording a user through a camera or a microphone found on a user's mobile device or any other camera or microphone found in a ubiquitous environment (3\%). In total 6.5\% of the studies on privacy address identity exposure in the context of privacy and security as two conflicting goals. A lesser amount of studies (3\%) focus their research on so-called over-privileged apps \cite{Wu2013}, which request permissions to device's resources and may lead not only to recording user behavior and tracking his/her whereabouts, but also to accessing personal photos and other information stored on or accessed through a user's mobile device. Although we only found studies about smartphones, the issue of over-privileged apps can also be extended to other devices that allow a user to install apps, such as tablet computers or wearable devices.

\subsection{Attacks}

The results of our SLR (see Figure \ref{fig:attacks_privacy}a) reveal that \textbf{DoS attacks} have been the most commonly addressed attacks over the last decade. This type of attack aims at making the resources and services unavailable to its intended users. For example, by jamming the network (25\% DoS attacks), exhausting the devices' resources (12\%), providing false trust ratings to nodes participating in a communication (9\%) or degrading the services (9\%).

The second most frequent type of attacks belong to the category of \textbf{masquerading or impersonation attacks} (20\% attacks in ubiquitous computing) which include, amongst other, man-in-the-middle (42\% of impersonation attacks), spoofing (28\%) and Sybil\footnote{In a Sybil attack a single entity impersonates multiple identities to gain a disproportionally large influence. For example, in an online voting or in a reputation/trust context (see, e.g. \cite{Wang13}).} (19\%) attacks.

\textbf{Eavesdropping attacks} are the third most commonly addressed attacks in ubiquitous computing, covering 16\% of attacks in ubiquitous computing. This type of an attack refers to a group of attacks in which a malicious user monitors communication to gain some knowledge about confidential information or interferes with the communication channel by modifying messages. The former one is known as passive eavesdropping. One example of such an attack is \textit{shoulder surfing} (11\% of eavesdropping attacks), in which an adversary observes the contents on a screen of a mobile device by secretly looking above the user's shoulder. Another example is snooping (15\%), which can be defined as the act of secretly observing what a person is typing on his/her computer. However, this attack can be performed in a more sophisticated manner by using software to monitor someone's activity on a computer. In another type of an eavesdropping attack, an adversary not only passively observes a user's sensitive information, but also tries to repeat, delay, alter, or intercept messages being transmitted. Such attacks belong to the group of \textit{active eavesdropping} (74\% of eavesdropping attacks) and include replay attacks\footnote{Note that this categorization of replay attacks under ``active eavesdropping'' follows the argument of the corresponding papers and assumes that an attacker first has to (actively) intercept network traffic in order to select and re-send certain messages afterwards.}. 
	
The fourth most commonly addressed attacks refer to \textbf{cryptanalytic attacks}, covering 11\% of the attacks, which include password cracking attacks (57\% of cryptanalytic attacks), side-channel attacks (25\%), such as electromagnetic attacks and acoustic cryptanalysis key search, and other cryptanalytic attacks, such as ciphertext, birthday, preimage, and key generation attacks.

An increasing number of devices used by users opens more ways for attacks. Nowadays it is not only possible to attack the user's bank account, but also cause a serious damage to a user's health by
modifying sensor readings from wearable medical devices (10\% of attacks in the category \textbf{Other attacks} or track user's location (6\% of attacks belonging to \textbf{geo-location inference attacks}). Apart from modifying sensor readings, attacks in the group \textit{Other} include smudge attacks described as a type of an attack in which an adversary tries to repeat the pattern by using the fingerprint smudges left on the device's screen, RFID-blocker attacks, zero day attacks, session hijacking, and physically stealing a mobile device.

\section{Defenses}
\label{sec:defenses}

\begin{figure}[!htbp]
	\centering
		\includegraphics[scale=0.7]{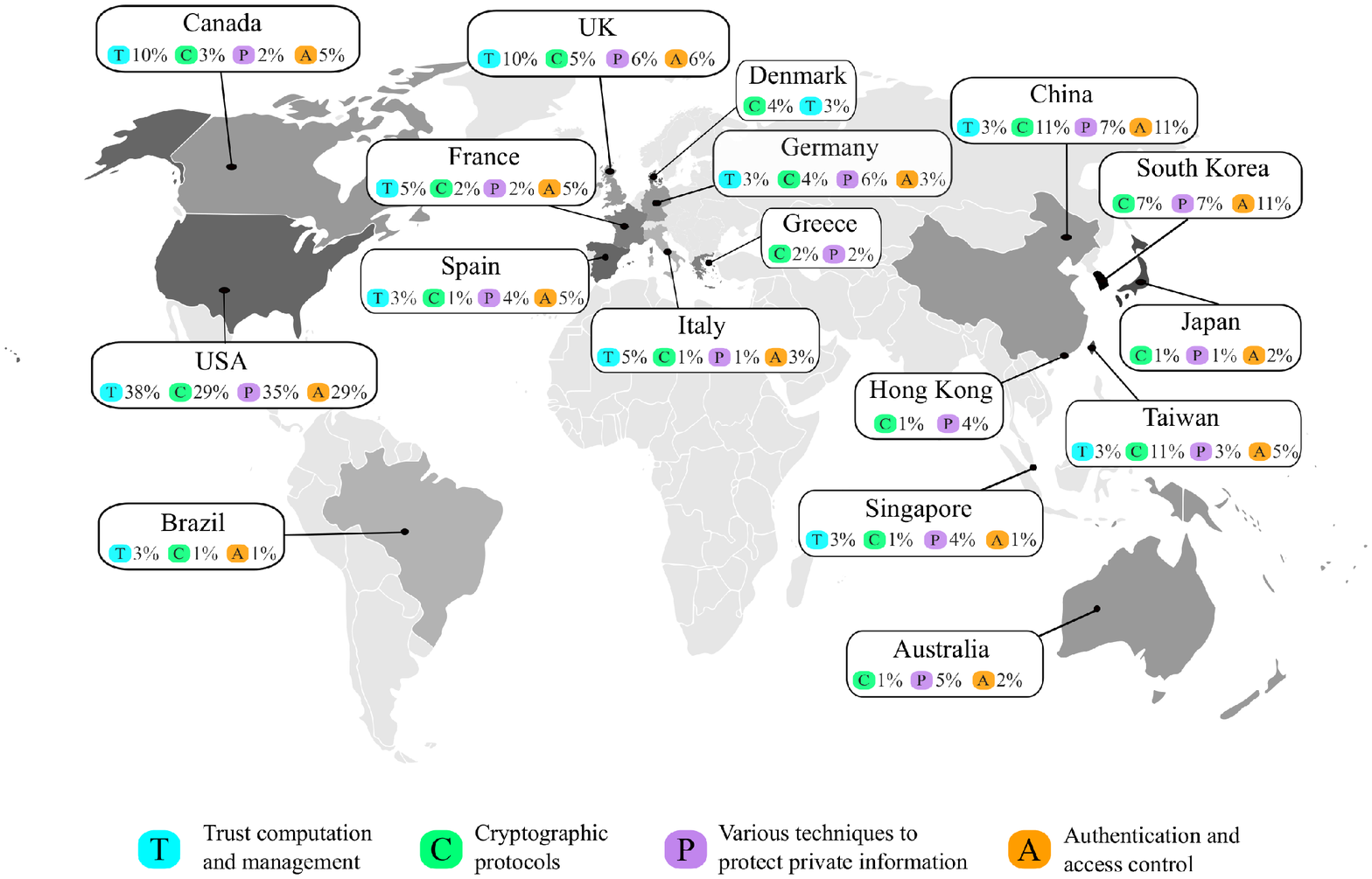}
	\caption{Four most commonly proposed groups of solutions found in the SLR.}
	\label{fig:countries_pic_solutions}
\end{figure}

The results of our SLR indicate that the research community most
frequently proposes the following four groups of defenses: (1) trust
computation and management, (2) cryptographic protocols, (3) design of
authentication and access control mechanisms, and (4) additional
mechanisms that aim at privacy protection. Figure
\ref{fig:countries_pic_solutions} presents percentages of the defense
mechanisms proposed by researchers from different regions.

\subsection{Trust computation and management}

\begin{figure}[ht]
	\centering
		\includegraphics[scale=0.7]{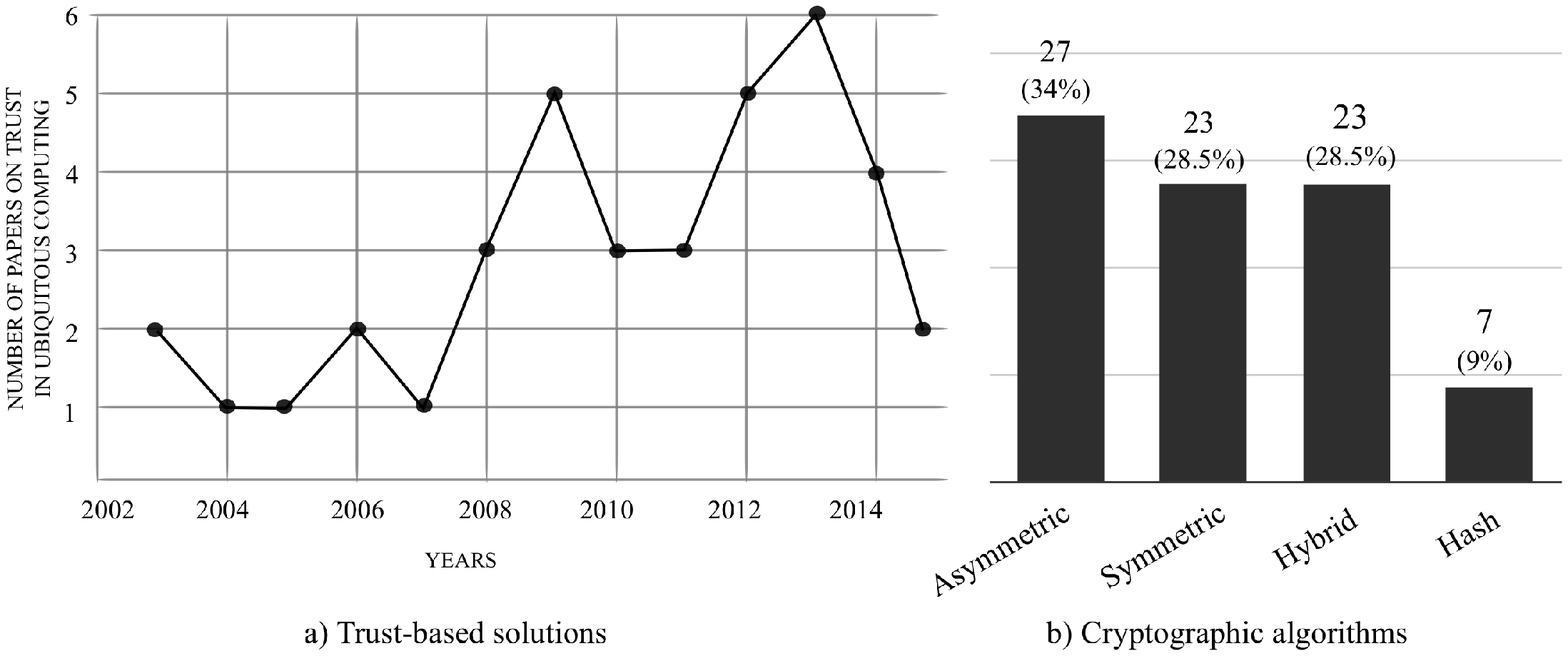}
	\caption{Number of trust-based solutions and cryptographic algorithms proposed over the last decade.}
	\label{fig:trust}
\end{figure}

In a ubiquitous computing network, mobile nodes can engage themselves
in a spontaneous interaction with other nodes. With the rising number
of mobile devices participating as network nodes, it becomes
challenging to ensure that all will behave properly. The importance of
designing effective trust models for such dynamic interactions and
fast-changing network topologies has gained significance over the last
decade (see Figure \ref{fig:trust}a). Typically, trust models rely on
the trust values assigned or derived based on the node's behavior.
Some models assume direct interaction with a newcomer node to evaluate
its behavior, while other approaches use recommendations from nodes
that have previously interacted with the newcomer node. If a node is
evaluated as well-behaved, it may continue communicating with other
nodes in the network. Malicious nodes are excluded from the network.
Since nodes may become non-cooperative over time (see,
e.g.\cite{Das12}), it is important to ensure dynamics in trust
computation by monitoring a node's actions and recomputing the assigned
trust values.

\subsection{Cryptographic protocols}

With the proliferation of small and resource-constrained devices,
lightweight cryptography \cite{Eisenbarth2007} has appeared as a new
branch of cryptography which differs from it's counterpart by a lesser
computational overhead. In the SLR, we have identified a considerable
amount of studies (63 (28\%)) that investigate the use of cryptographic
algorithms, out of which 56\% report on a lightweight protocol
design. We have also investigated the types of cryptographic
algorithms. As seen in Figure \ref{fig:trust}b), 
protocols in ubiquitous computing mostly rely on three types of
cryptographic algorithms, 30\% on symmetric and 29\% asymmetric, while a
majority of studies proposes a combination of both (hybrid approaches
are proposed in 36.5\% studies).

\subsection{Authentication and access control}

While analyzing the studies for our SLR, we have
identified a number of requirements that need to be ensured while
designing authentication and access control mechanisms for ubiquitous
computing environments:

\begin{enumerate}
\item \textbf{Dynamics} (22 (28\%) studies on authentication and access control). Dynamics can be divided
  into three distinct categories.
	\begin{enumerate}
        \item \textbf{Adaptivity to context (context-awareness).} In a
          traditional distributed environment, access to a computer
          system and sensitive data is often restricted based on the
          roles assigned to its users. This approach is called
          role-based access control (RBAC). However, ubiquitous
          computing assumes that users are able to move freely and
          request, receive, and use services at any given time. Such
          requirements pose a need for a context-centric access
          control mechanism that can dynamically determine contexts of
          users by at least considering the two dimensions of time
          (temporal dimension) and space (spatial dimension) (see, e.g.
          \cite{Strembeck2004}).
        \item \textbf{Adaptivity to user's behavior.} A user may
          request a service by using one of the devices he/she owns.
          For example, a user may use his/her e-book reader,
          smartphone, or a tablet to access his/her collection of
          purchased e-books. With a rising number of devices
          participating in a ubiquitous computing environment, it is
          important to ensure that only the well-behaved entities are
          granted access rights. Therefore, maintaining trust
          relationships and assigning attributes to users based on
          their behavior or managing trust relationships are important
          tasks.
        \item \textbf{Dynamic recognition of users.} An ongoing
          discussion about the opposing goals of privacy and security
          is also present in ubiquitous computing (addressed in
          6.5\% of the studies). While
          permitting well-behaved users to access a certain service,
          they should be given a possibility not to disclose their
          personal information, such as their name and location.
          However, anonymity opens the door for malicious users to
          access services without a permission. Our SLR identified two
          types of solutions: identity-based
          (43 (84\%) studies on
          authentication mechanisms) and non-identity based
          (3 (6\%) studies).  The idea
          behind the former lies in entity recognition, which is based
          on standard identifiers, such as a person's name,
          credentials, or biometrics, which includes capturing
          biometric features (e.g.\ iris, fingerprint, palm) and
          recognizing a user's individual activity (e.g.\ walking
          patterns, heartbeat). A smaller number of solutions propose
          a non-identity-based authentication mechanism which uses a
          user's trustworthiness (instead of a user's identity) to
          authenticate him/her to a service provider. Computation of
          trustworthiness is based on the anonymous user's reputation
          levels.
        \end{enumerate}
      \item \textbf{Non-intrusiveness (13 (16.5\%).}
        Traditional authentication mechanisms require users to
        interact with their devices by typing in their username or
        password. Yet, in order to realize a vision of ubiquitous
        computing, it is important to ensure \textit{unobtrusiveness}
        of such security mechanisms. One way to realize this vision is
        to use magnetic cards (see, e.g. \cite{Jaspher2012}) and other
        physical authentication tokens, which are already a widespread
        mechanism used in, for example, skiing resorts. However,
        problems occur when users leave their personal token behind or
        when it gets stolen. To address this problem, researchers
        suggest replacing such tokens by wearable devices that may
        allow for a continuous, unobtrusive authentication (see, e.g.
        \cite{sun2008}).
      \item \textbf{Speed (1 (1\%)).} While designing
        authentication schemes it is important to avoid latency during
        the handoff (\textit{handover}) process, which occurs each time
          a node changes its access point.
      \item \textbf{Lightweight authentication protocol (1 (1\%)).} Due to the resource constraints of small devices,
        authentication protocols are designed so that they require
        less computation overload.
\end{enumerate}

\subsection{Privacy protection mechanisms}

Using context-aware services, users with mobile and wearable devices
can automatically collect, sense, share, and process information.  Our
SLR has identified a number of privacy preserving mechanisms summarized
in Table \ref{tab:PrivacyProtectionMechanisms}.

\begin{table}[ht]
\scriptsize
	\centering
		\begin{tabularx}{\textwidth}{p{4cm} p{9.3cm} p{1.5cm} }
		\toprule
			\textbf{Mechanism type} & \textbf{Description} & \textbf{Nr./\%}\\
		\midrule	
			Masking mechanism & Hiding identities by replacing a real ID by an anonymous ID from the untrusted parties.  & 18 (29\%)\\
			Privacy protection layer for mobile apps & Security analysis of apps to find undesirable properties in security configuration and a lightweight certification of apps at install-time. Use of policies that regulate assignment of permissions. & 9 (15\%)\\
			Proximity detection scheme & Trust computation is based on the \textit{encounter}, which assumes two persons being in a close proximity for a period of time. If there is a mutual interest to establish communication, two users establish a trust relationship only if they can convince each other that they encountered each other at some time in the past. Absence of mutual interest or proof of the encounter prohibits users to share personal information, i.e. they remain anonymous to each other. & 8 (13\%) \\
			Game-based approach & A privacy-preserving mechanism can be modeled as a competition game, which enables a designer to find the optimal location-privacy protection mechanism. An example of such a model is a Bayesian Stackelberg competition game in which the user plays first by choosing a protection mechanism and commits to it by running it on his/her actual location. The adversary plays next by estimating the user’s location, but with incomplete information about the user's true location. Another approach proposed in the studies is a \textit{credits-earning game}, in which a user earns points by contributing data without leaking which data has been contributed. & 6 (10\%) \\
			Consent and notification & A system that takes notification and consent while recording users. & 6 (10\%) \\
			Negotiation approach & This approach aims to find proper information to be exposed by allowing users to negotiate with the services on submitting data elements according to their privacy preferences. & 5 (8\%) \\ 
			Other & Other techniques encompass recommenders for privacy sharing, tag identification schemes, and other that could not be grouped into any of the proposed categories. & 4 (7\%)\\
			Obfuscation & Deliberate degradation of the quality of information. & 3 (5\%)\\
			Blocker tags for RFID-based systems & Use of a passive RFID tag that can simulate many RFID tags simultaneously. & 2 (3\%)\\
			\bottomrule
		\end{tabularx}
	\caption{Privacy protection mechanisms.}
	\label{tab:PrivacyProtectionMechanisms}
\end{table}

\section{Countries participating in the research}

\begin{figure}[h]
\centering
		\includegraphics[scale=0.5]{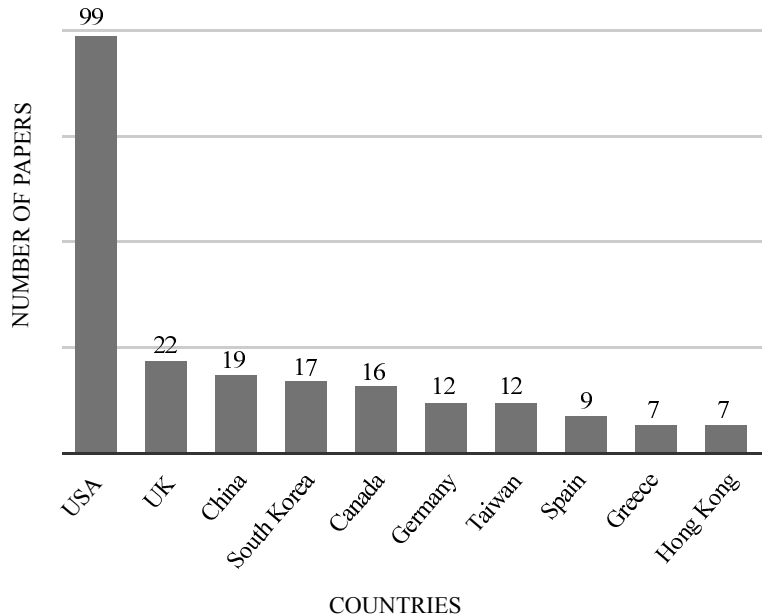}
	\caption{Top 10 countries participating in the research on security in ubiquitous computing.}
	\label{fig:countries}
\end{figure} 

Figure \ref{fig:countries_pic_solutions} provides an overview of the
countries which predominantly contributed to
  the most commonly addressed defense mechanisms identified in our
  SLR. Figure \ref{fig:countries} provides a list of the ``top 10''
  countries based on the number of papers identified in our SLR.
  According to our findings, there is a noticeable predominance of the
  USA, Western Europe, and Eastern Asia.

\section{Conclusion}

With the rise of the ubiquitous computing era, we are facing a variety
of threats which aim at exploiting sensitive information or corrupting
user data. As ubiquitous technologies and
  devices have found their way into users' homes and often accompany
  their owners throughout the day, it has become important more than
  ever to design defense mechanisms that protect a user from
  ill-taught spying, destroying device's resources, or even
  endangering a person's health by modifying medical sensor readings.
  Over the last decade, researchers have been demonstrating attacks
  novel to ubiquitous computing and pointing to a number of
  vulnerabilities of this dynamic environment. The results of our SLR
  indicate that ubiquitous computing, along with the contextual
  information that is collected and processed by mobile devices
  together with the heterogeneity of the participating devices, opens
  novel requirements while designing security mechanisms. As we expect
  devices to continue getting smaller and more powerful, research
  endeavors will have to face the challenge of designing unobtrusive
  and lightweight security solutions that are able to dynamically
  adapt to their environment as well as their user's behavior.

\end{document}